\documentclass[dvips]{appolb}
\usepackage[dvips]{graphicx}
\usepackage{graphicx}
\usepackage{latexsym}

\graphicspath{{plots/}}
\DeclareGraphicsExtensions{.eps.gz,.eps,.ps,.ps.gz}
\newcommand{\lwig}{\mbox{\,\raisebox{.3ex}
    {$<$}$\!\!\!\!\!$\raisebox{-.9ex}{$\sim$}\,}}
\newcommand{\gwig}{\mbox{\,\raisebox{.3ex}
    {$>$}$\!\!\!\!\!$\raisebox{-.9ex}{$\sim$}}\,}
\newcommand{\rav}{\langle\rho\rangle}
\newcommand{\x}{x_{\rm Bj}}

\newcommand{\vr}{\mathbf{r}}
\newcommand{\rb}[1]{\raisebox{1.5ex}[-1.5ex]{#1}}
\def\funp{{I\!\!P}}
\def\Journal#1#2#3#4{{#1} {\bf #2} (#4) #3}

\def\NPA{{ Nucl. Phys.} A}
\def\NPB{{ Nucl. Phys.} B}
\def\PLB{{ Phys. Lett.}  B}
\def\PRL{ Phys. Rev. Lett.}
\def\PRD{{ Phys. Rev.} D}
\def\PR{Phys. Rev.}
\def\ZPC{{ Z. Phys.} C}

\def\CPC{ Comput. Phys. Commun.}

\def\JPG{ J. Phys. G }

\begin{document}
\title{
\vspace{-2.0cm}{\normalsize\rightline{DESY
02-093}\rightline{hep-ph/0207052}}
\vspace{1cm}
\bf Instantons and Saturation in the Colour Dipole Picture
\thanks{Presented at the Xth International Workshop on Deep Inelastic Scattering (DIS~2002), Cracow, 30 April - 4 May 2002}
}
\author{F.~Schrempp and A.~Utermann
\address{
DESY, Notkestra{\ss}e 85, D-22603 Hamburg, Germany}
}

\maketitle

\begin{abstract}
We pursue the intriguing possibility that
larger-size instantons build up  diffractive
scattering, with the marked instanton-size scale $\rav\approx 0.5$ 
fm being reflected in the conspicuous ``geometrization'' of soft QCD.
As an explicit illustration,  
the known instanton contribution to DIS is transformed into the
intuitive colour dipole picture. With the help of lattice results,  
the $q\bar{q}$-dipole size $r$ is carefully increased towards hadronic
dimensions. Unlike pQCD, one now observes a competition between two
crucial length scales: the dipole size $r$ and the size $\rho$ of the
background instanton that is sharply localized around $\rav\approx
0.5$ fm. For $r\,\gwig\,\rav$, the dipole cross section indeed
saturates towards a geometrical limit, proportional to the area
$\pi\,\rav^2$, subtended by the instanton. 
\end{abstract}
\begin{figure} 
\begin{center}
\parbox{3.5cm}{\includegraphics*[width=3.5cm]{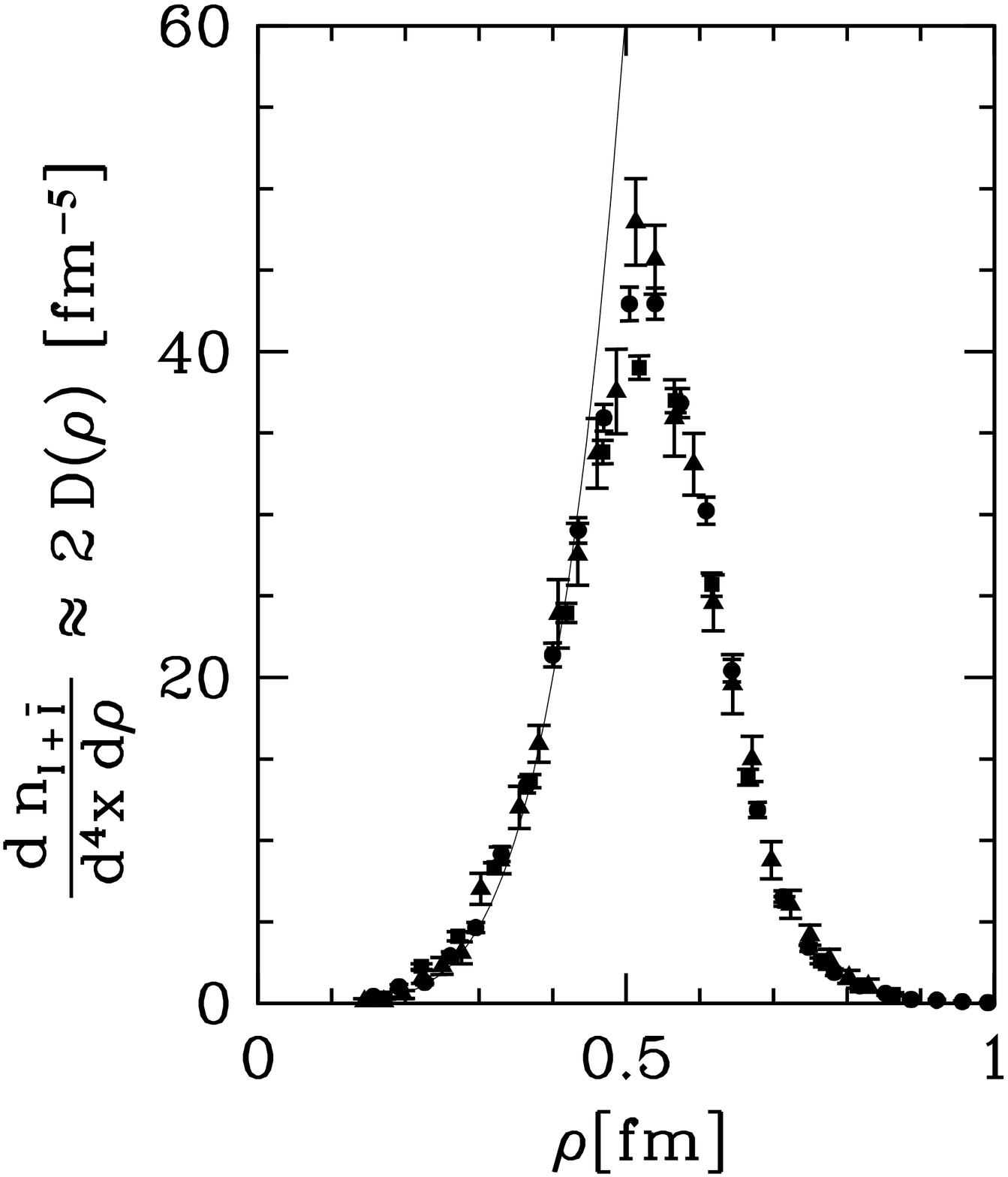}}\hfill
\parbox{9cm}{\includegraphics*[width=9cm]{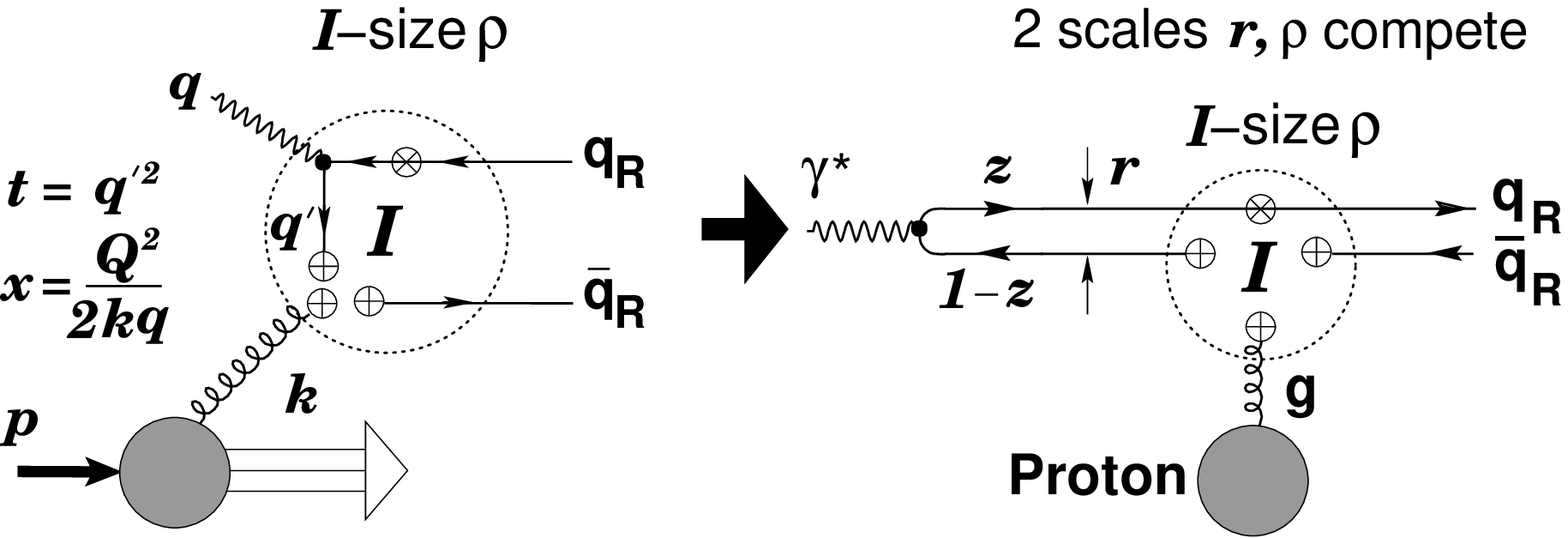}}
 \caption[dum]{\small\label{pic} (Left) $I+\overline{I}$-size
 distribution from the 
lattice~\cite{ukqcd,rs-lat}. Both the sharply defined $I$-size scale
$\langle\rho\rangle \approx 0.5$ fm and the parameter-free agreement with
$I$-perturbation theory for $\rho\lwig 0.35$ fm are apparent.
(Right) Transscription of the simplest
 \mbox{$I$-induced} process ($n_f = 1,\ n_g = 0$)  with variables $x$ and $t$
 into the colour dipole picture with the variables $z$ and $\mathbf r$}
\end{center}
\end{figure}

\vspace{1ex}
QCD instantons~\cite{bpst} are non-perturbative
fluctuations of the gluon fields, with a size distribution {\em sharply}
localized around $\rav\approx 0.5~{\rm fm}$ according to lattice
simulations~\cite{ukqcd} (Fig. \ref{pic} left). They are well known to induce,
chirality-violating processes, absent in conventional perturbation
theory~\cite{th}.  Deep-inelastic
scattering (DIS) at HERA has been shown to offer a unique
opportunity~\cite{rs1} for discovering such processes
induced by {\it small} instantons ($I$) through a sizeable
rate~\cite{mrs,rs2,rs-lat} and a characteristic final-state
signature~\cite{rs1,qcdins,rs3}.  
The intriguing but non-conclusive excess of events, found recently in the first
dedicated search for instanton-induced processes in DIS at
HERA~\cite{h1-final}, has also been reported at this meeting.

The validity of $I$-perturbation
theory in DIS is warranted by some (generic) hard momentum scale
$\mathcal{Q}$ that ensures a dynamical
suppression~\cite{mrs} of contributions from larger size instantons
with $\rho\gwig \mathcal{O}(1/\mathcal{Q})$. Here, the above mentioned
intrinsic instanton-size scale $\rav\approx 0.5~{\rm fm}$ is correspondingly
unimportant.    

This paper, in contrast, is devoted to the intriguing question about
the r{\^o}le of {\em larger-size} instantons and the associated intrinsic scale
\mbox{$\rav\approx 0.5~{\rm fm}$}, for decreasing ($Q^2,\ \x$)
towards the soft regime. We shall briefly report on a detailed study~\cite{us2}
of the interesting possibility that larger-size instantons may well be
associated with a dominant part of soft high-energy scattering, or even make up
diffractive scattering
altogether~\cite{levin,shuryak1,fs,shuryak2}. We shall argue below that the 
intrinsic instanton scale $\rav$ is reflected in the conspicuous {\it
geometrization} of soft QCD. 

There are two immediate qualitative reasons for this idea. 

First of all, instantons represent truely non-perturbative gluons that
naturally bring in an intrinsic size scale $\rav\approx 0.5~{\rm fm}$ 
of hadronic dimension (Fig.~\ref{pic} left). The instanton-size scale
happens to be surprisingly close to a corresponding ``diffractive''-size
scale, $R_\funp =\,R\,\sqrt{\alpha^\prime_\funp/\alpha^\prime}
\approx 0.5$ fm, resulting from simple dimensional rescaling along
with a generic hadronic size \mbox{$R\approx 1$ fm} and the abnormally small $\funp$omeron slope
$\alpha^\prime_\funp\approx \frac{1}{4}\,\alpha^\prime$ in terms of the normal,
universal Regge slope  $\alpha^\prime$. 

Secondly, we know already from $I$-perturbation theory that the instanton
contribution tends to strongly increase towards the infrared
regime~\cite{bb,rs1,rs2,qcdins}. The mechanism for the decreasing
instanton suppression with increasing energy is known since a long
time~\cite{sphal2,shuryak2}: Feeding increasing energy into the scattering
process makes the picture shift from one 
of tunneling between vacua ($E\approx 0$) to that of the actual
creation of the sphaleron configuration~\cite{sphal1} on top of the potential
barrier of height~\cite{rs1} $E = M_{\rm
sphaleron}\propto\frac{1}{\alpha_s\rho_{\rm eff.}}$. In a second step,
the action is real and the sphaleron then decays into a multi-parton
final state.  
  
The familiar colour dipole picture~\cite{dipole} represents a
convenient and intuitive framework for investigating the transition from
hard to soft physics (diffraction) in DIS at small $\x$. 
%
At the same time, this picture is very well suited for studying the crucial
interplay between the $q\bar{q}$-dipole size $r$ 
and the instanton size $\rho$ in an explicit and well-defined manner, as we
shall summarize next. The details may be found in Ref.~\cite{us2}.

The large difference of the $\gamma^\ast\rightarrow
q\overline{q}$-dipole formation and 
$(q\overline{q})$-$P$ interaction times in the proton's rest frame at
small $\x$ is at the root of the familiar factorized expression of
the inclusive photon-proton cross sections, 
\begin{equation}
\sigma_{L,T}(\x,Q^2) 
=\int_0^1 d z \int d^2\vr\; |\Psi_{L,T}(z,r)|^2\,\sigma_{\rm dipole}(r,\ldots), 
\label{dipole-cross}
\end{equation}
in terms of the modulus squared of the 
(light-cone) wave function of the virtual photon, calculable in pQCD
($\hat{Q}=\sqrt{z(1-z)} Q;\ r=\mid{\mathbf r}\mid$), 
\begin{equation}
\label{wavefu}
 \mid\Psi_{ L,T}(z,r)\mid^{\,2}
 =e_q^2\frac{6\alpha}{4\pi^2} N_{L,T}(z)\,\hat{Q}^{\,2}\,{\rm
 K}_{0,1}(\hat{Q}r)^{\,2};\
 \begin{array}{l}N_L=4z(1-z)\\N_T=z^2+(1-z)^2\end{array}\end{equation} 
and the dipole\,-$P$  cross section $\sigma_{\rm
dipole}(r,\ldots)$. The variables in Eq.~(\ref{dipole-cross}) denote
the transverse $(q\overline{q})$-size $\mathbf r $ 
and the photon's longitudinal momentum fraction $z$ carried by the quark. 
$\Psi_{L,T}(z,r)$ contains the
dependence on the $\gamma^\ast$-helicity. Moreover, one derives
~\cite{dipole,dipole-pqcd} and expects, respectively,
$$
\sigma_{\rm dipole}\ \left\{\begin{array}{lrcl}
\sim & \pi\,r^{\,2},&
{r}^2\lwig\mathcal{O}(\frac{1}{Q^2}),&\mbox{\rm\ ``colour
transparency''~\cite{dipole,dipole-pqcd}},\\[1ex] 
\approx &{\rm constant},&r\,\gwig\, 0.5 {\rm\ fm},& \mbox{\rm\
``hadron-like,\ saturation''.}\end{array}\right.
$$ 
The strategy is now to transform the known results on
$I$-induced processes in DIS  into this intuitive colour dipole
picture. Here, for reasons of space, we
restrict the discussion to the most transparent case of the simplest
$I$-induced process~\cite{mrs},
$\gamma^\ast\,g\Rightarrow q_{\rm R}\
\overline{q}_{\rm R}$, for one flavour  and no final-state
gluons (Fig.~\ref{pic} right). The more realistic case with gluons and
three light flavours, using the $I\overline{I}$-valley
approximation, may be found in Ref.~\cite{us2}.

The idea is to consider first large $Q^2$ and appropriate cuts on the
variables $z$ and $r$, such
that \mbox{$I$-per}turbation theory holds. By exploiting the lattice results
on the instanton-size distribution (Fig.~\ref{pic} left),
we shall then carefully increase the $q\bar{q}$-dipole size $r$ towards
hadronic dimensions. 


Let us start by recalling the results from Ref.~\cite{mrs},  
\vspace{-1ex}
\begin{eqnarray}
\sigma_{L,T}(\x,Q^2)&=&
\int_{\x}^1 \frac{d x}{x}\left(\frac{\x}{ x}\right)G\left(\frac{\x}{x},\mu^2\right)\int d  t \frac{d \hat{\sigma}_{L,T}^{\gamma^* g}(x,t,Q^2)}{d t};\,\\
\frac{d\hat{\sigma}_{L}^{\gamma^* g}}{d  t}&=&\frac{\pi^7}{2} \frac{e_q^2}{Q^2}\frac{\alpha}{\alpha_s}\left[x(1-x) \sqrt{t u}\,
\frac{R(- t)-R(Q^2)}{t+Q^2}-(t\leftrightarrow  u)\right]^{\,2}
\label{mrs}
\end{eqnarray}
and a similar expression for $d\hat{\sigma}_{T}^{\gamma^*
g}/d\,t$. 

Eqs.~(\ref{mrs}) involve the master integral $R(\mathcal{Q})$ with dimensions of a length, 

\begin{equation}
R(\mathcal{Q})=\int_0^{\infty} d\rho\;D(\rho)\rho^5(\mathcal{Q}\rho)\mbox{K}_1(\mathcal{Q}\rho).
\label{masterI}
\end{equation}

The $I$-size distribution $D(\rho)$ enters in Eq.~(\ref{masterI}) as a
crucial building block of the 
$I$-calculus. For small $\rho$ (probed at large $\mathcal{Q}$)
$D(\rho)$ is calculable within $I$-perturbation theory
\cite{th}. For larger $I$-size $\rho$ (as relevant for smaller 
$\mathcal{Q}$) $D(\rho)$ is known from lattice simulations
(Fig.~\ref{pic} left). A striking feature is the strong peaking, whence 
$R(0)=\int_0^{\infty} d\rho\;D_{\rm lattice}(\rho)\rho^5\approx\, \rav$.

With an appropriate change of variables (Fig.~\ref{pic} right)
and a $2d$-Fourier 
transformation, Eqs.~(\ref{mrs}) may indeed be cast into a
colour dipole form, 
\vspace{-1ex}
\begin{equation}
\sigma_{L,T}=
\int_{\x}^1 \frac{d x}{x}
\int d t\, \frac{d \hat{\sigma}_{L,T}^{\gamma^* g}}{d
t}\,\{\ldots\}\Rightarrow \int dz\int d^2\mathbf{r} \;
\left(|\Psi_{L,T}|^2 \sigma_{\rm dipole}\right)^{(I)}. 
\end{equation}
Like in pQCD-calculations~\cite{dipole-pqcd}, we
invoke the familiar ``leading-$\log(1/\x)$'' - approximation, $\x/x
G(\x/x,\mu^2) \approx \x G(\x,\mu^2)$.  
In terms of the familiar pQCD wave function (\ref{wavefu}) of the
photon, we then obtain e.\,g., 
\begin{eqnarray}
&& \left(\left|\Psi_L\right|^2\sigma_{\rm dipole}\right)^{(I)}
 \approx\, \mid\Psi_L^{\rm pQCD}(z,r)\mid^{\,2}\,
\frac{1}{\alpha_s}\,\x\, G(\x,\mu^2)\,\frac{\pi^8}{12} \label{result}\\[1ex]
&&\times\left(\int_0^\infty\,d\rho D(\rho)\,\rho^5\,\left\{\frac{-\frac{d}{dr^2}\left(2 r^2
\frac{\mbox{K}_1(\hat{Q}\sqrt{r^2+\rho^2/z})}{\hat{Q}\sqrt{r^2+\rho^2/z}}\right)}{{\rm K}_0(\hat{Q}r)}-(z\leftrightarrow 1-z) \right\} \nonumber
\right)^2.
\end{eqnarray}
As expected, one explicity observes a 
{\em competition} between two crucial length scales in 
Eq.~(\ref{result}): the size $r$ of the 
$q\bar{q}$-dipole and the typical size of the background
instanton of about $\rav\approx 0.5~{\rm fm}$. 
Like in pQCD, the {\it asymmetric} configuration, $z \gg
1-z$ or $1-z \gg z$, obviously dominates.

The validity of strict $I$-perturbation theory, $D(\rho)= D_{I-{\rm
pert}}(\rho)$ in Eq.\,(\ref{masterI}), requires the presence of a hard
scale $\mathcal{Q}$ along with certain cuts. However,  
after replacing $D(\rho)$ by $D_{\rm lattice}(\rho)$ (Fig.~\ref{pic} left),
these restrictions are at least no longer  
necessary for reasons of convergence of the $\rho$-integral
(\ref{masterI}) etc.,  and one may tentatively 
increase the dipole size $r$ towards hadronic dimensions. 

 Next, we note in Eq.~(\ref{result}), 
\begin{equation}
-\frac{d}{d\,r^2}\left(2\,r^2\frac{{\rm K}_1\left(\hat{Q}\sqrt{r^2+\rho^2/z}\right)}{\hat{Q}\sqrt{r^2+\rho^2/z}}\right) \approx \left\{\begin{array}{rl}
-\frac{{\rm K}_1\left(Q\, \rho\sqrt{1-z}\right)}{Q\,\rho\sqrt{1-z}}&\frac{r^2\,z}{\rho^2}\Rightarrow 0,\\[2ex]
{\rm K}_0\left(\hat{Q}\,r\right)&\frac{r^2\,z}{\rho^2}\mbox{\ large}.  
\end{array}\right.
\label{approx}
\end{equation}
Due to the strong peaking of $D_{\rm
lattice}(\rho)$ around \mbox{$\rho\approx\rav$}, one finds from 
Eqs.~(\ref{result}, \ref{approx}) ($z\gg 1-z$ without restriction) for
the limiting cases of interest,
\vspace{-2ex}
\begin{equation}
\begin{array}{c|c}
r&\rule[-2mm]{0mm}{7mm}\left(\mid\Psi_{L,T}\mid^{\,2} \sigma_{\rm dipole}\right)^{(I)}\\[1ex]\hline
\rule[2mm]{0mm}{3mm}
r^2 \Rightarrow 0&\mathcal{O}(1),\mbox{\rm \ but\ exponentially\ small\
for\ large}\ \hat{Q},\\[1ex]
 &\rule[-2mm]{0mm}{7mm} \mid\Psi^{\rm \,pQCD}_{L,T}\mid^{\,2}\, \sigma_{\rm
 dipole}^{(I)}\hspace{2ex}\mbox{\rm\ with}\\[2ex]
\rb{$r^2 \gwig \rav^2$ :}&\rule[-3mm]{0mm}{3mm}\sigma^{(I)}_{\rm
dipole}=\frac{1}{\alpha_s}\,\x\,G(\x,\mu^2)\,\frac{\pi^8}{12}\,\left(\int_0^\infty\,d\rho\,D_{\rm lattice}(\rho)\,\rho^5\right)^2.  
\end{array}
\label{final}
\end{equation}
{\em In conclusion:} As apparent in Eq.~(\ref{final}), the dipole
cross section indeed \mbox{\em saturates} for large $r^2/\rho^2\approx
r^2/\rav^2$  towards a {\em geometrical limit}, proportional to the area
$\pi\,R(0)^2=\pi\left(\int_0^\infty\,d\rho\,D_{\rm
lattice}(\rho)\,\rho^5\right)^2$, subtended by the instanton. Clearly,
without the crucial information about $D(\rho)$ from the 
lattice \mbox{(Fig.~\ref{pic} left)}, the result would be infinite. Note the
inverse power of $\alpha_s$ in front of $\sigma_{\rm  dipole}^{(I)}$
in Eq.~(\ref{final}), signalling its non-perturbative nature.

\end{document}